\documentclass[twocolumn, secnumarabic, amssymb, nobibnotes, aps, prb, superscriptaddress]{revtex4-2}

\usepackage{graphicx,amsmath}

\begin{document}

\title{
Comparative $^{181}$Ta-NQR Study of Weyl Monopnictides TaAs and TaP: Relevance of Weyl Fermion Excitations
}

\author{Tetsuro~Kubo}%
\affiliation{Max Plank Institute for Chemical Physics of Solids, 01187 Dresden, Germany}
\affiliation{Department of Physics, Okayama University of Science, Okayama 700-0005, Japan}

\author{Hiroshi~Yasuoka}%
\affiliation{Max Plank Institute for Chemical Physics of Solids, 01187 Dresden, Germany}

\author{Bal\'{a}zs~D\'{o}ra}%
\affiliation{Department of Theoretical Physics and MTA-BME Lend\"{u}let Topology and Correlation Research Group, Budapest University of Technology and Economics, Budapest 1521, Hungary}

\author{Deepa~Kasinathan}%
\altaffiliation[Present address: ]{MHP Management- und IT-Beratung GmbH. Film- und Medienzentrum K\"onigsallee 49, 71638 Ludwigsburg, Germany}
\affiliation{Max Plank Institute for Chemical Physics of Solids, 01187 Dresden, Germany}

\author{Yurii~Prots}%
\affiliation{Max Plank Institute for Chemical Physics of Solids, 01187 Dresden, Germany}

\author{Helge~Rosner}%
\affiliation{Max Plank Institute for Chemical Physics of Solids, 01187 Dresden, Germany}

\author{Takuto~Fujii}%
\altaffiliation[Present address: ]{University of Hyogo, Graduate School of Material Science, Hyogo 678-1297, Japan}
\affiliation{Max Plank Institute for Chemical Physics of Solids, 01187 Dresden, Germany}

\author{Marcus~Schmidt}%
\affiliation{Max Plank Institute for Chemical Physics of Solids, 01187 Dresden, Germany}

\author{Michael~Baenitz}%
\email{michael.baenitz@cpfs.mpg.de}
\affiliation{Max Plank Institute for Chemical Physics of Solids, 01187 Dresden, Germany}

\date{
\today
}%

\begin{abstract}
Based on our first detailed $^{181}$Ta nuclear quadrupole resonance (NQR) studies from 2017 on the Weyl semimetal TaP, we now extended our NQR studies to another Ta-based monopnictide TaAs. In the present work, we have determined the temperature-dependent $^{181}$Ta-NQR spectra, the spin-lattice relaxation time $T_{1}$, and the spin-spin relaxation time $T_{2}$. We found the following characteristic features that showed great contrast to what was found in TaP: (1) The quadrupole coupling constant and asymmetry parameter of EFG, extracted from three NQR frequencies, have a strong temperature dependence above $\sim$80\,K that cannot be explained by the density functional theory calculation incorporating the thermal expansion of the lattice. (2) The temperature dependence of the spin-lattice relaxation rate, $1/T_{1} T$, shows a $T^{4}$ power law behavior above $\sim$30\,K. This is a great contrast with the $1/T_{1} T \propto T^{2}$ behavior found in TaP, which was ascribed to the magnetic excitations at the Weyl nodes with a temperature-dependent orbital hyperfine coupling. (3) Regarding the nuclear spin-spin interaction, we found the spin-echo signal decays with the pulse separation simply by a Lorentzian function in TaAs, but we have observed spin-echo modulations in TaP that is most likely due to the indirect nuclear spin-spin coupling via virtually excited Weyl fermions. From our experimental findings, we conclude that the present NQR results do not show dominant contributions from Weyl fermion excitations in TaAs.
\end{abstract}

\maketitle

\section{Introduction}
It is well known by now that the topological properties in materials open a new world in condensed matter physics. Particularly, topological semimetals, such as Dirac, Weyl, or line-node semimetals, are gapless states of matter characterized by their nodal band structures and surface states \cite{Franz}. Weyl semimetals are realized in systems without spatial-inversion or time-reversal symmetry even more with the strong spin-obit coupling. Quite interesting new phenomena such as ultra-high mobility \cite{mobility2015}, surface Fermi arcs \cite{SFA}, and chiral magnetic effect \cite{CME} are expected to emerge in these materials. Furthermore, in topological semimetals, we found new types of quasiparticles, Dirac- and Weyl-fermions, and those excitations have exhibited fascinating properties that have been the subject of many theoretical and experimental investigations. The first target materials in this field were taken among the monopnictides $TMPn$ ($TM$ = Nb, Ta; $Pn$ = P, As). Visualization of the nodal structure in topological semimetals has been realized by ARPES (angle-resolved photoemission spectroscopy) in the topologically protected surface states \cite{TPSS}. Indirectly, the large negative magnetoresistance \cite{CANMR}, optical and resistivity measurements \cite{optic}, and chiral anomalies \cite{ABJ2016} are believed to be associated with those quasiparticles. In addition to the surface-sensitive probes like ARPES and electron spin resonance (ESR), the microscopic measurements which enable us to study the static and dynamical properties of quasiparticles as a bulk are highly expected. Along this line, we have succeeded for the first time to explore the Weyl fermion excitations in TaP through the temperature dependence of nuclear quadrupole resonance (NQR) relaxation rate, $1/T_{1} T$ \cite{TaPNQR}. There, we have demonstrated that in addition to the $T^{4}$ power law of $1/T_{1} T$ associated with the linear dispersion of Weyl nodes near the Fermi level, we have pointed out the importance of fluctuations in Dirac/Weyl-type orbital currents to the relaxation channel through the characteristic temperature dependence of the orbital hyperfine interaction \cite{Aorb2016}. This scenario is supported by the theory explicitly and the overall temperature dependence of $1/T_{1} T \propto T^{2}$ has been interpreted properly \cite{Aorb2019}.

In this paper we present an extended study of the sister compound, TaAs, using the same $^{181}$Ta-NQR technique. TaAs has been claimed to be a typical example of the Weyl nodal semimetal from band structure calculations. Regarding the nodal structure, both compounds have Weyl points near the Fermi level, $E_{\rm F}$, 14\,meV below $E_{\rm F}$ for the W2 Weyl points in TaAs \cite{TaAsQO}, while 13\,meV above $E_{\rm F}$ in TaP \cite{TaPQO}.

In the following, we will first briefly describe the experimental technique, then discuss the temperature dependence of NQR parameters, $\nu_{\rm Q}$ and $\eta$. This will be followed by the temperature dependence of nuclear magnetic relaxation time, $T_{1}$, and a comparison of spin-echo decay curves of TaAs and TaP, along with their interpretations.

\section{Experimental}
Basically, we followed the experimental procedure of previous $^{181}$Ta-NQR experiments and analysis \cite{TaPNQR}. Here, we briefly describe the essence of them.

Samples used in the present NQR experiments were prepared by the chemical transport reaction (CTR) method. Starting from microcrystalline powder synthesized by reacting 3-nine Tantalum and 6-nine Arsenic, single crystals of TaAs were grown in a temperature gradient from 900\,$^\circ$C (source) to 1000\,$^\circ$C (sink), and a transport agent concentration of 13\,mg/cm$^{3}$ iodine. The crystals obtained by the CTR were characterized by electron-probe-microanalysis and powder X-ray diffraction (XRD) to ensure the single phase, tetragonal $I 4_{1} m d$ (\#109) structure.

Temperature-dependent powder XRD was performed at the beamline ID22 at the European Synchrotron Research Facility (ESRF) in Grenoble in a temperature range between 80 and 300\,K with a wavelength $\lambda = 0.39997$\,\AA.

The NQR experiments were mostly carried out with high-quality polycrystals prepared by powdering several single crystals. The NQR spectra and the nuclear magnetic relaxation times were measured using a standard pulsed (spin-echo) NMR apparatus (Apollo, TecMag). The $^{181}$Ta-NQR spectra were taken using the frequency sweep method under zero applied magnetic field. In order to avoid any artificial broadening, fast Fourier-transformed (FFT) spin-echo signals were summed across the spectrum (FFT-summation), or the real part of spin-echoes was integrated after appropriate phase adjustments.

\begin{figure}[htbp]
 \centering
 \includegraphics[width=1.00\linewidth]{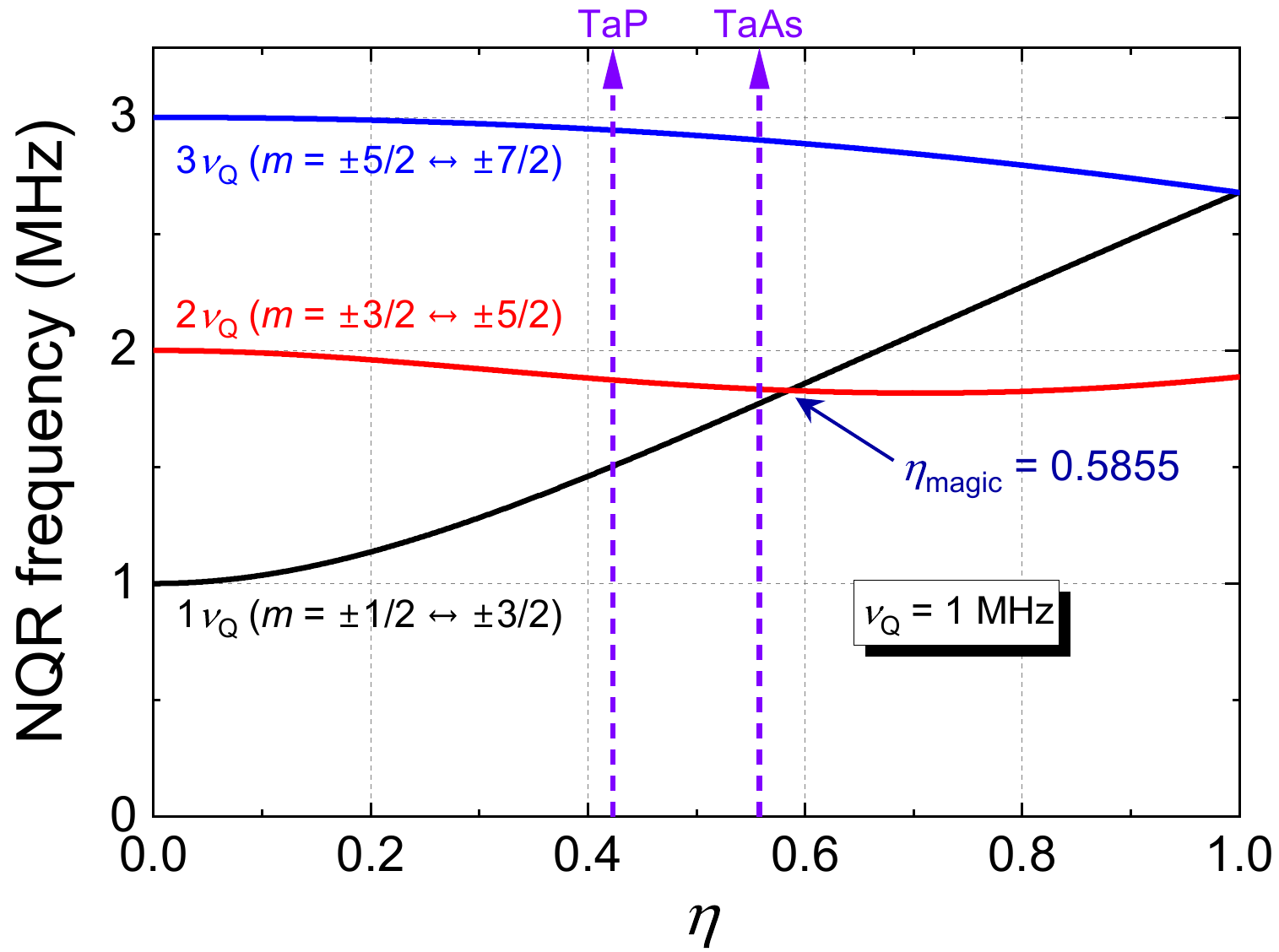}
 \caption{(Color online) EFG asymmetry parameter, $\eta$, dependence of NQR frequencies for $I = 7/2$. Here, the quadrupole coupling parameter, $\nu_{\rm Q}$, is set as 1 MHz. The cross point between $1 \nu_{\rm Q}$ and $2 \nu_{\rm Q}$ lines at $\eta = 0.5855$ is called as the ``magic eta''. $\eta$ values determined from the experimental NQR frequencies for TaAs ($\eta = 0.558$) and TaP ($\eta = 0.423$) are shown by dashed arrows.}
 \label{fig:EFG}
\end{figure}

The quadrupole Hamiltonian can be written, using a set of principal axes \cite{Slichter}, as
\begin{equation}
 \mathcal{H}_{\rm Q} = \frac{e^{2} q Q}{4 I (2 I - 1)} \left[3 I_{Z}^{2} - I (I + 1) + \frac{\eta}{2} (I_{+} - I_{-}^{2})\right],
 \label{eqn:hq}
\end{equation}
where $e q$ is the largest component of the electric field gradient (EFG) tensor, $V_{ZZ}$, and $e Q$ the nuclear quadrupole moment. The EFG tensor is generally defined as $|V_{XX}| \leq |V_{YY}| \leq |V_{ZZ}|$ with the asymmetry parameter, $\eta \equiv (V_{XX} - V_{YY})/V_{ZZ}$. The quadrupole-split nuclear energy levels, $E_{m}$, and the resultant transition frequencies can be readily calculated numerically by diagonalizing Eq.\,(\ref{eqn:hq}). For $\eta = 0$, the energy levels can simply be expressed as,
\begin{equation}
 E_{m} = \frac{1}{6} h \nu_{\rm Q} \left[3 m^{2} - I (I + 1)\right], ~ \nu_{\rm Q} = \frac{3 e^{2} q Q}{2 I (2 I - 1) h},
\end{equation}
where $\nu_{\rm Q}$ is the quadrupole coupling constant. The NQR occurs for the transition between two levels $m$ and $m + 1$, and the resonance condition can be written as $f_{\rm Q} = \nu_{\rm Q} (2 |m| + 1)/2$. Hence, three NQR lines for $^{181}$Ta with $I = 7/2$ are expected at $1 \nu_{\rm Q}$, $2 \nu_{\rm Q}$, and $3 \nu_{\rm Q}$ with equal spacing. For the finite $\eta$ values, the calculated NQR frequencies with $\nu_{\rm Q} = 1$ MHz for respective transitions are shown in Fig.\,\ref{fig:EFG}.

Since the nuclear spin-lattice (longitudinal) relaxation time, $T_{1}$, is extremely long in TaAs at low temperatures (typically several hundred seconds), we mostly employed the progressive saturation method \cite{Progr} to measure the recovery of nuclear magnetization below 130\,K, as was in the case of TaP \cite{TaPNQR}. Above 130\,K, a conventional inversion recovery method is employed. At 130\,K, both methods yield essentially the same $T_{1}$ value.

The recovery of nuclear magnetization was fitted to the theoretical function for the magnetic relaxation in NQR lines for $^{181}$Ta ($I = 7/2$) nucleus with finite $\eta = 0.558$ \cite{etaT1},
\begin{align}
 M_{n}(t) = & M_{0} [1 - \{Q_{1} \exp(-3.03 t/T_{1}) \nonumber \\
 & + \{Q_{2} \exp(-8.260 t/T_{1}) \nonumber \\
 & + \{Q_{3} \exp(-17.074 t/T_{1})\}],
 \label{eqn:recovery}
\end{align}
where $Q_{n}$ are constants depending on which NQR transition is excited. For the preset $T_{1}$ measurements, we typically used 2$\nu_{\rm Q}$-line corresponding to the $\pm 3/2 \leftrightarrow \pm 5/2$ nuclear quadrupole transition. In this case, $Q_{1}$, $Q_{2}$ and $Q_{3}$ are 0.076, 0.021, and 0.903, respectively.

\begin{figure}[htbp]
 \centering
 \includegraphics[width=1.00\linewidth]{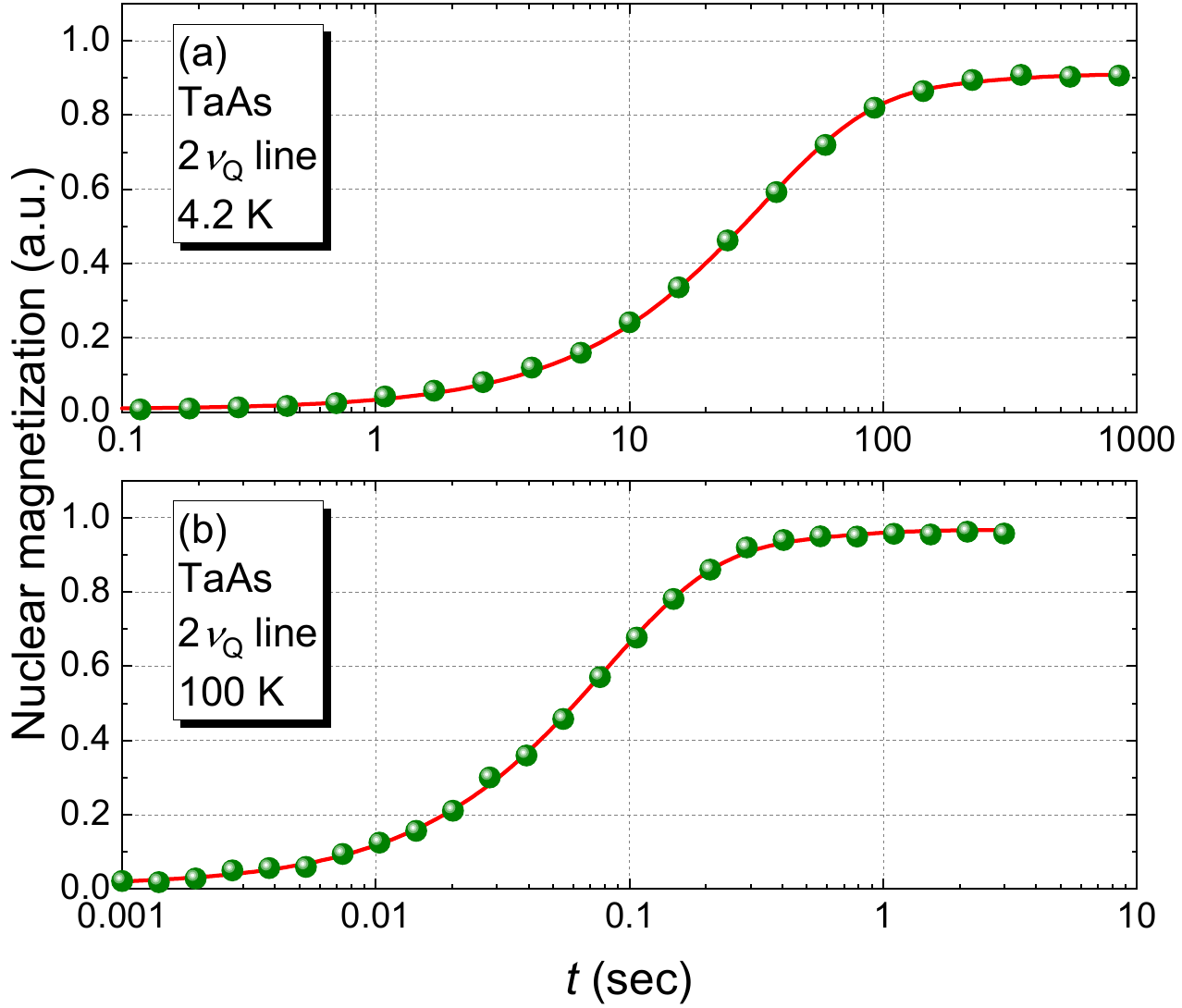}
 \caption{(Color online) The recovery curves of nuclear magnetization measured by the spin-echo amplitude for the $2 \nu_{\rm Q}$ line at (a) 4.2 and (b) 100\,K. Solid lines are the least-squares fit to Eq.\,(\ref{eqn:recovery}) for magnetic fluctuations. For details, see text.}
 \label{fig:recovery}
\end{figure}

Figure \ref{fig:recovery} shows the recovery of nuclear magnetization measured by the progressive saturation method at (a) 4.2 K and (b) 100 K for the $^{181}$Ta-NQR $2 \nu_{\rm Q}$-line in TaAs. Solid lines are the least-squares fitting to Eq. (\ref{eqn:recovery}). For both temperatures, experimental data are perfectly fitted by the theoretical curve, verifying that the nuclear relaxation is governed by magnetic fluctuations as in the case of TaP \cite{TaPNQR}.

For the measurements of the nuclear spin-spin (transverse) relaxation time, $T_{2}$, we simply measure spin-echo amplitude, $E$, as a function of the time between the first exciting and the second refocusing pulses. The repetition time between spin-echo sequences was taken to be sufficiently longer than $T_{1}$ (typically 8--10 times longer than $T_{1}$ value) to avoid a saturation effect.

In order to extract the EFG theoretically, we performed band structure calculations using the density functional theory (DFT) solid-state code FPLO \cite{FPLO}. We used the Perdew-Wang parametrization of the local density approximation (LDA) for the exchange-correlation functional \cite{PW,wannier90}. The strong spin-orbit coupling in TaAs is taken into account by performing full-relativistic calculations, wherein the Dirac Hamiltonian with a general potential is solved. The quadrupole coupling $\nu_{\rm Q}$ can be obtained by the calculated EFG at the Ta nuclear site which is defined as the second partial derivative of the electrostatic potential $V(\vec{r})$ at the position of the nucleus $V_{ij} = (\partial_{i} \partial_{j} V(0) - \delta_{ij} \Delta V(0)/3)$.

\section{Experimental Results and Discussion}
In this section, we present the static and dynamic properties revealed by the temperature dependence of the EFG parameters and nuclear magnetic relaxation times with phenomenological discussions.

\subsection{NQR spectra and their temperature dependence}
\begin{figure}[htbp]
 \centering
 \includegraphics[width=1.00\linewidth]{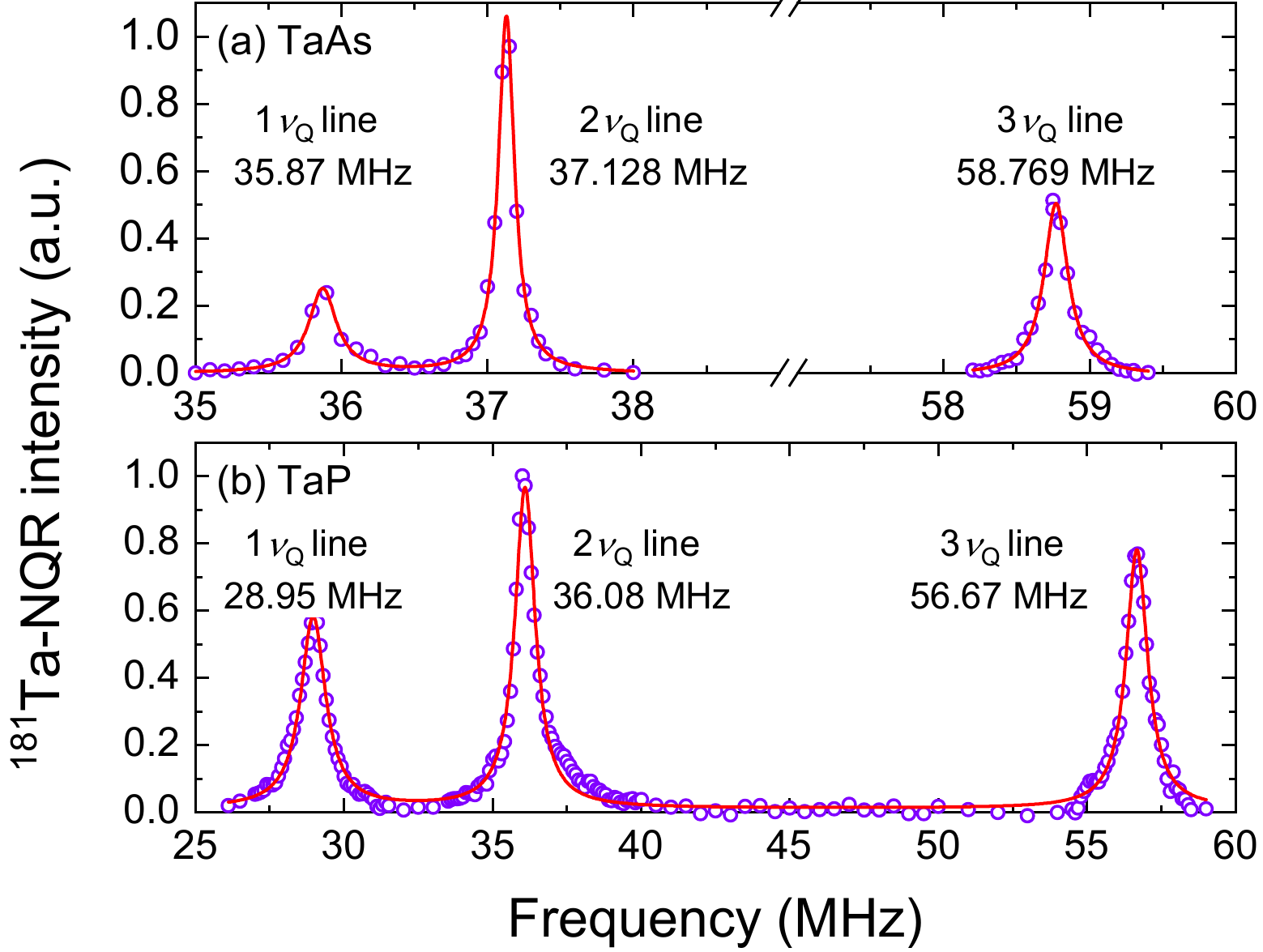}
 \caption{(Color online) A $^{181}$Ta-NQR spectrum in (a) TaAs in comparison with that in (b) TaP. Both spectra were taken at 4.2\,K and curves are Lorentzian fit of the line profiles.}
 \label{fig:spectrum}
\end{figure}

A typical example of $^{181}$Ta-NQR spectra in TaAs measured at 4.2\,K is shown in Fig.\,\ref{fig:spectrum}(a) for three NQR transitions. From the lowest frequency we define the lines as $1 \nu_{\rm Q}$ ($\pm 1/2 \leftrightarrow \pm 3/2$), $2 \nu_{\rm Q}$ ($\pm 3/2 \leftrightarrow \pm 5/2$) and $3 \nu_{\rm Q}$ ($\pm 5/2 \leftrightarrow \pm 7/2$) lines. For comparison, we also depict a similar spectrum of TaP in Fig.\,\ref{fig:spectrum}(b). It is immediately seen that the frequency difference between $1 \nu_{\rm Q}$ and $2 \nu_{\rm Q}$ lines is smaller for TaAs than TaP. It means that the $\eta$ value is larger for TaAs than TaP as indicated in Fig.\,\ref{fig:EFG}. The EFG parameters, $\nu_{\rm Q}$ and $\eta$, were calculated using the same manner described in Ref. \cite{TaPNQR} and shown in Table \ref{tab:DFT}, together with values for TaP and NbP. For NbP, $\nu_{\rm Q}$ and $\eta$ are extracted from the single-crystal $^{93}$Nb-NMR spectrum at $\sim$6.3\,T, 4.2\,K. The calculated values of $\nu_{\rm Q}$ agrees with the experimental values of all compound within 4\%.

\begin{figure}[htbp]
 \centering
 \includegraphics[width=1.00\linewidth]{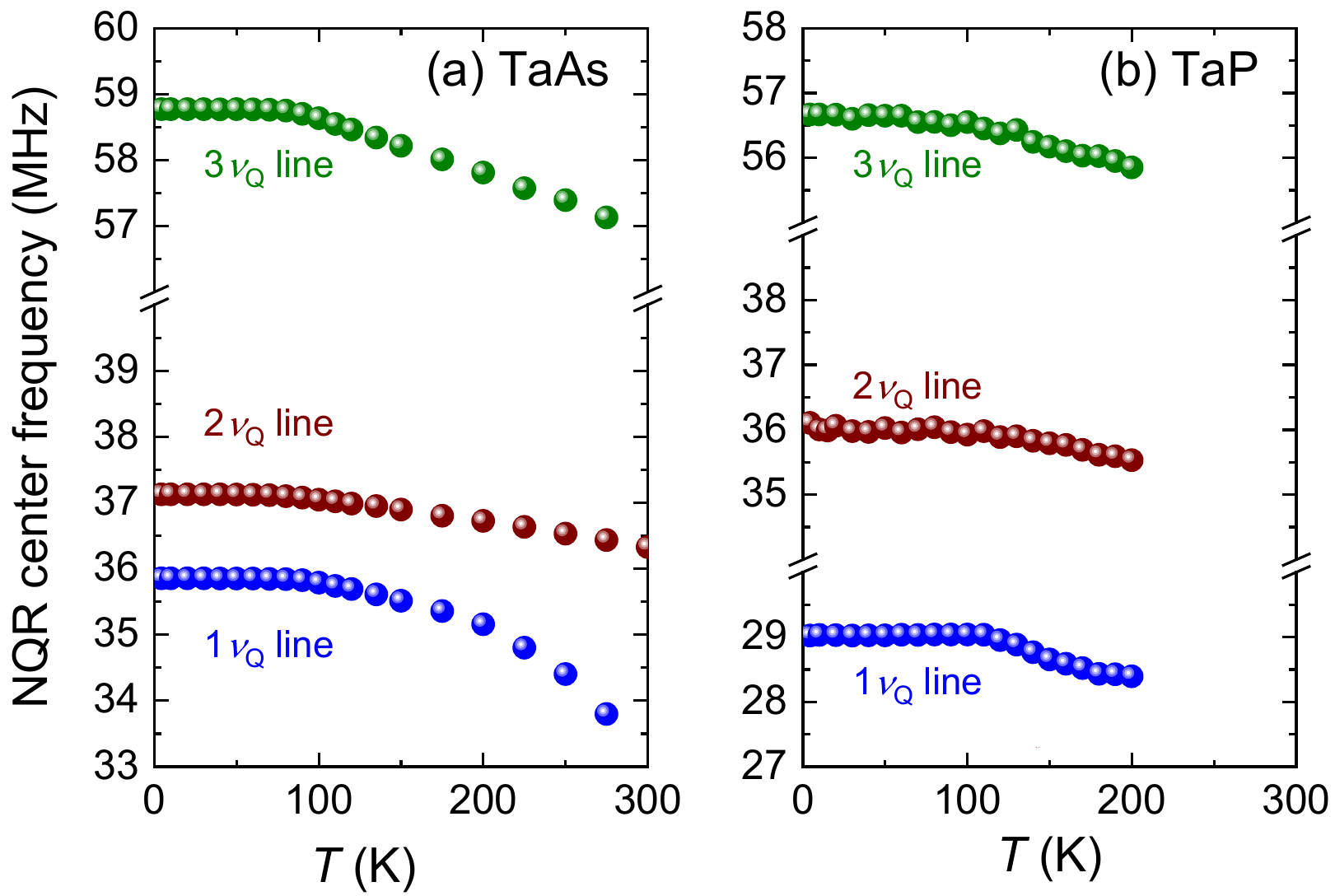}
 \caption{(Color online) Temperature dependence of the NQR peak frequencies in (a) TaAs in comparison with those in (b) TaP. Here $1 \nu_{\rm Q}$, $2 \nu_{\rm Q}$, and $3 \nu_{\rm Q}$ line correspond to $\pm 1/2 \leftrightarrow \pm 3/2$, $\pm 3/2 \leftrightarrow \pm 5/2$, and $\pm 5/2 \leftrightarrow \pm 7/2$ quadrupole transitions, respectively.}
 \label{fig:peak}
\end{figure}

\begin{figure}[htbp]
 \centering
 \includegraphics[width=1.00\linewidth]{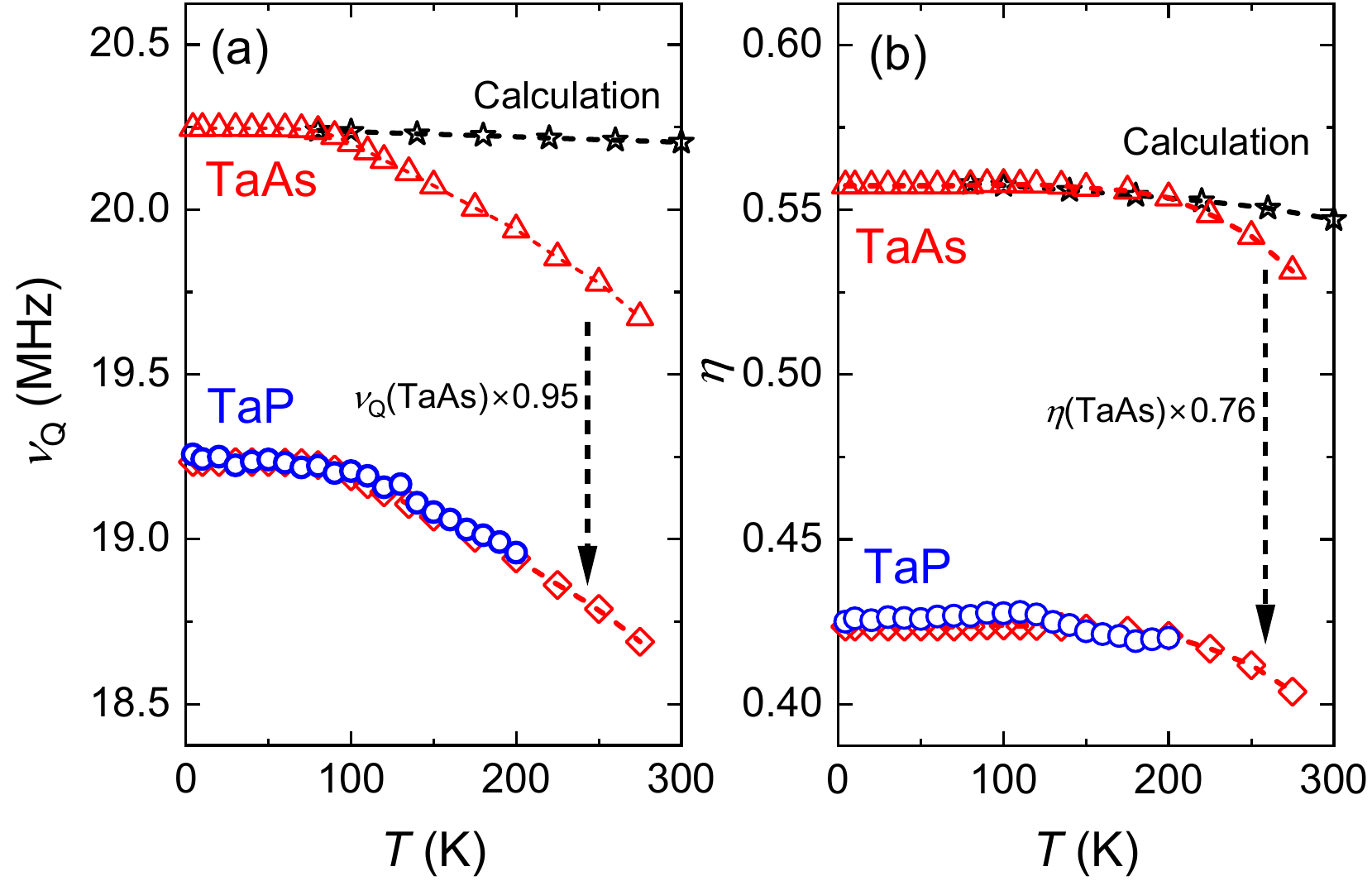}
 \caption{(Color online) Temperature dependences of quadrupole coupling parameter, $\nu_{\rm Q}$, and asymmetry parameter of the EFG, $\eta$, for TaAs and TaP are shown in panels (a) and (b), respectively. Both values were extracted from experimental data shown in Fig.\,\ref{fig:peak} by diagonalizing Eq.\,(\ref{eqn:hq}). The dashed curves with open stars are DFT calculated values of $\nu_{\rm Q}$ and $\eta$ using the lattice parameters of respective temperatures shown in Table \ref{tab:lattice}. The calculated $\nu_{\rm Q} (T)$ follows the empirical form with $\nu_{{\rm Q}}(T) = \nu_{{\rm Q} 0} (1 - \alpha T^{3/2})$, $\nu_{{\rm Q} 0} = 20.24$\,MHz, $\alpha = 3.92 \times 10^{-7}$\,K$^{-3/2}$.}
 \label{fig:nuQeta}
\end{figure}

The temperature dependence of NQR frequencies is plotted in Fig.\,\ref{fig:peak}. From the observed NQR frequencies we can extract $\nu_{\rm Q}$ and $\eta$ by diagonalizing Eq.\,(\ref{eqn:hq}) and those temperature dependences, $\nu_{\rm Q} (T)$ and $\eta (T)$, are shown in Fig.\,\ref{fig:nuQeta}(a) and (b) for TaAs (open triangles) and TaP (open circles), respectively. In general, $\nu_{\rm Q}$ is expected to decrease with increasing temperature due to the thermal expansion of the lattice, and is often discussed by an empirical formula, $\nu_{\rm Q} (T) = \nu_{{\rm Q}0} (1 - \alpha T^{3/2})$. Actually, $\nu_{\rm Q} (T)$ is well fitted to this empirical formula below $\sim$100\,K for both TaAs and TaP, but the experimental data fall more rapidly above $\sim$100\,K.

\begin{table}
 \caption{\label{tab:DFT}DFT calculated EFG parameters $\nu_{\rm Q}$ and $\eta$. The values of $\nu_{\rm Q}$ and $\eta$ are in good agreement with those extracted from the NQR frequencies in TaAs and TaP \cite{TaPNQR}. Also, those obtained by the quadrupole split $^{93}$Nb-NMR spectrum in a single-crystalline NbP are in good agreement with the DFT calculation.}
 \begin{tabular}{ccccc} \hline
   & DFT calculation & & Experimental & \\
   & $\nu_{\rm Q}$ (MHz) & $\eta$ & $\nu_{\rm Q}$ (MHz) & $\eta$ \\ \hline
   TaAs & 20.830 & 0.625 & 20.249 & 0.558 \\
   TaP & 20.057 & 0.330 & 19.250 & 0.423 \\
   NbP & 0.667 & 0.1912 & 0.65 & 0.15 \\ \hline
 \end{tabular}
\end{table}

In particular, for TaAs, we have made the DFT calculation for EFG parameters by putting the measured lattice parameters measured by synchrotron XRD at selected temperatures above 80\,K shown in Table \ref{tab:lattice}. As the temperature decreases, the lattice parameters exhibit a slight, monotonous reduction while maintaining a constant $c/a$ ratio, indicative of isotropic behavior. The calculated $\nu_{\rm Q} (T)$ and $\eta$ are shown by stars in Fig.\,\ref{fig:nuQeta}(a). As can be seen from the figure, the thermal expansion of the lattice cannot account for the temperature dependence of the observed NQR parameters.

It should be noted here that the fractional decrease of $\nu_{\rm Q} (T)$ and $\eta (T)$ is very similar as shown by scaling TaAs results (shown by open open diamonds) to TaP by factors of 0.95 and 0.76 for $\nu_{\rm Q} (T)$ and $\eta (T)$, respectively.

\begin{table}
 \caption{\label{tab:lattice}Lattice parameters measured by synchrotron XRD for TaAs at selected temperatures above 80\,K.}
 \begin{tabular}{ccccc}
  \hline
  $T(K)$ & $a$ (\AA) & $c$ (\AA) & $c/a$ & $V$ (\AA$^{3})$ \\ \hline
  300 & 3.43752 & 11.64762 & 3.38838 & 137.63 \\
  260 & 3.43642 & 11.64374 & 3.38833 & 137.50 \\
  220 & 3.43563 & 11.64089 & 3.38828 & 137.40 \\
  180 & 3.43492 & 11.63826 & 3.38822 & 137.32 \\
  140 & 3.43437 & 11.63627 & 3.38818 & 137.25 \\
  100 & 3.43375 & 11.63397 & 3.38812 & 137.17 \\
   80 & 3.43349 & 11.63299 & 3.38809 & 137.14 \\ \hline
 \end{tabular}
 \label{tab:lattice}
\end{table}

EFG generally has contributions from lattice symmetry and asymmetrical charge distribution around the nucleus in concern,
\begin{equation}
 e q (T) = e q_{\rm lattice} (T) + e q_{\rm el} (T).
\end{equation}
The first term was literally calculated by measured thermal lattice expansion and agrees well with the experimental values below 80\,K. On the other hand, experimental data deviate from the first term, showing an almost linear dependence on temperature above 80\,K. This fact suggests that the second term becomes dominant above 80\,K, indicating an unusual electronic contribution to the EFG is induced by a reason that is yet to be identified.

\subsection{Nuclear spin-lattice relaxation}
\begin{figure}[htbp]
 \centering
 \includegraphics[width=1.00\linewidth]{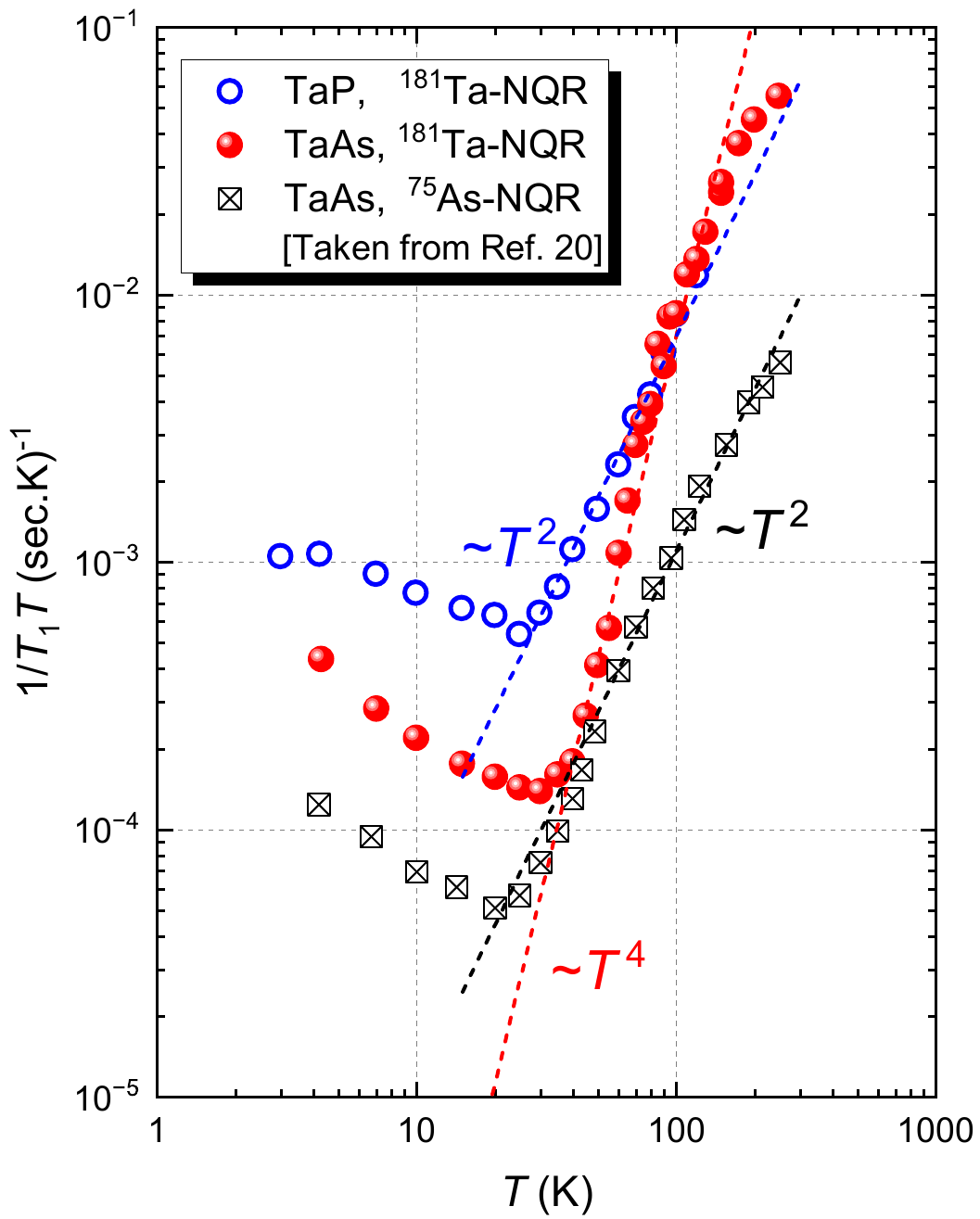}
 \caption{(Color online) Temperature dependence of $1/T_{1} T$ measured for the $2 \nu_{\rm Q}$ line in TaAs is shown by filled circles. For comparison, the similar data taken by $^{75}$As-NQR are shown in cross squares \cite{TaAsZheng}. Also, data in TaP are shown in open circles where the $T^{2}$ temperature dependence of $1/T_{1} T$ which is characteristic to the Weyl fermion excitations is seen above $\sim$30\,K \cite{TaPNQR}.}
 \label{fig:T1T}
\end{figure}

The temperature dependence of $^{181}$Ta nuclear spin-lattice relaxation rate divided by $T$, $1/T_{1} T$, in TaAs is shown in Fig.\,\ref{fig:T1T} together with previous data of TaP \cite{TaPNQR}. Also, we reproduced data of $^{75}$As-NQR taken by Wang {\it et al.} \cite{TaAsZheng}. There is a general tendency for the $1/T_{1} T$ exhibiting that from high-temperature power law type relaxation process crosses over to temperature-dependent Korringa type around $T^{*} \sim 20$--40\,K.

Quite generally, $1/T_{1} T$ can be expressed by using the wave vector ($q$) and frequency ($\omega$) dependent magnetic susceptibility, $\chi (q, \omega)$, characterizing the magnetic excitations in a system as,
\begin{equation}
 \frac{1}{T_{1} T} = \frac{2 \gamma_{\rm N}^{2} k_{\rm B}}{g^{2} \mu_{\rm B}^{2}} \sum_{q} A_{q}^2 \frac{\chi_{\perp}'' (q, \omega_{\rm N})}{\omega_{\rm N}},
\end{equation}
where $\gamma_{\rm N}$ is the nuclear gyromagnetic ratio, $k_{\rm B}$ the Boltzmann constant, $g$ the electron $g$-factor, $\mu_{\rm B}$ the Bohr magneton, $A_{q}$ the $q$-dependent hyperfine coupling constant, $\chi_\perp'' (q, \omega)$ the transverse component of imaginary part of $\chi (q, \omega)$, and $\omega_{\rm N}$ the NQR frequency. Since at present we do not have any plausible microscopic theory to calculate $\chi (q, \omega)$ in multiband systems like TaAs, we have adopted the theoretical $1/T_{1} T$ for non-interacting itinerant electrons based on the band structure calculation with random phase approximation (RPA). Also, since we do not have the information about $A_{q}$, we cannot perform a quantitative analysis of $1/T_{1} T$. So, we try to interpret the data qualitatively only using the shape of temperature dependence. In what follows we will discuss it by setting three cases: [Case-1] simple calculation from the density of states (DOS), [Case-2] in-gap states near the Fermi level, and [Case-3] Weyl fermion excitations.

\subsubsection*{Case-1: Simple $1/T_{1} T (T)$ from DOS}
\begin{figure}[htbp]
 \centering
 \includegraphics[width=1.00\linewidth]{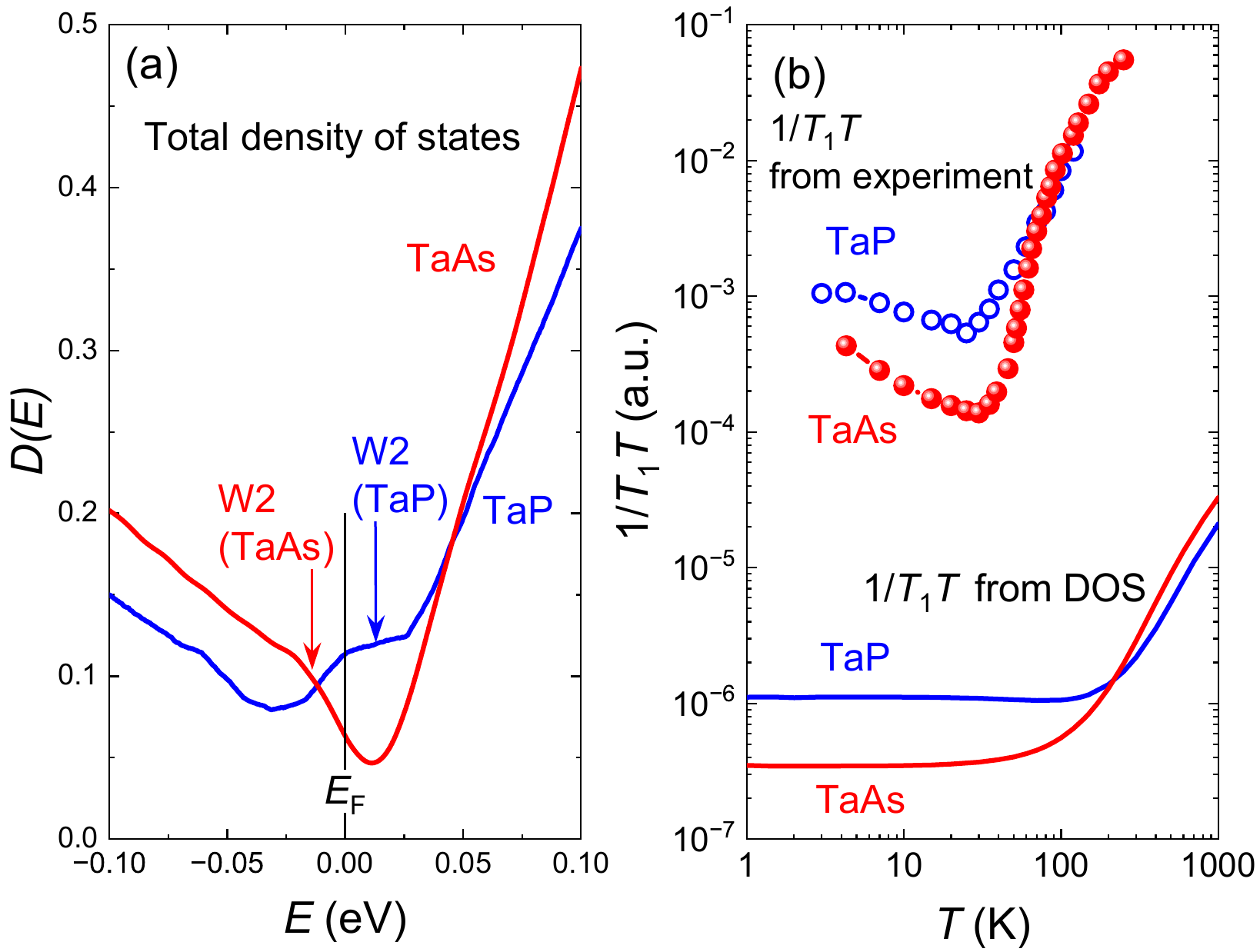}
 \caption{(Color online) (a) The calculated $D(E)$ curves within $\Delta E \sim \pm 1000$ K based on the band structure are shown for TaAs and TaP. (b) Temperature dependence of $1/T_{1} T$ calculated from DOS (solid lines) are compared with the experimental data for TaAs (filled circles) and TaP (open circles). Here, the hyperfine coupling constant $A_{\rm hf}$ is set to one and is assumed to be temperature-independent.}
 \label{fig:T1TDOS}
\end{figure}

Here, we have simply adopted the theoretical $1/T_{1} T$ for non-interacting itinerant electrons based on the band structure calculation with RPA. For such a system $1/T_{1} T$ may be expressed using density of state near the Fermi level as \cite{Slichter,Abragam},
\begin{equation}
 \frac{1}{T_{1} T} \propto \frac{A_{\rm hf}^{2}}{T} \int f (E - \mu_{\rm c}) [1 - f (E - \mu_{\rm c})] {D(E)}^{2} {\rm d} E,
\end{equation}
where $f (E)$ is a fermi distribution function, $D (E)$ the energy dependent DOS, and $\mu_{\rm c}$ the temperature-dependent chemical potential. If $A_{\rm hf}$ does not change with temperature and is set to one, the calculation of $1/T_{1} T$ is straightforward from calculated $D(E)$ based on the band structure shown as Fig.\,\ref{fig:T1TDOS}(a) for TaAs and TaP. The calculated $1/T_{1} T$ is shown by curved lines in Fig.\,\ref{fig:T1TDOS}(b). Also, we show the experimental $1/T_{1} T$ data of TaAs and TaP. We can clearly observe that, aside from the absolute value of $1/T_{1} T$, the temperature dependencies for experimental and calculated $1/T_{1} T$ do not match each other. This shows that the simple Korringa-type relaxation process cannot account for the experimental results observed.

\subsubsection*{Case-2: $1/T_{1} T$ from in-gap states}
\begin{figure}[htbp]
 \centering
 \includegraphics[width=1.00\linewidth]{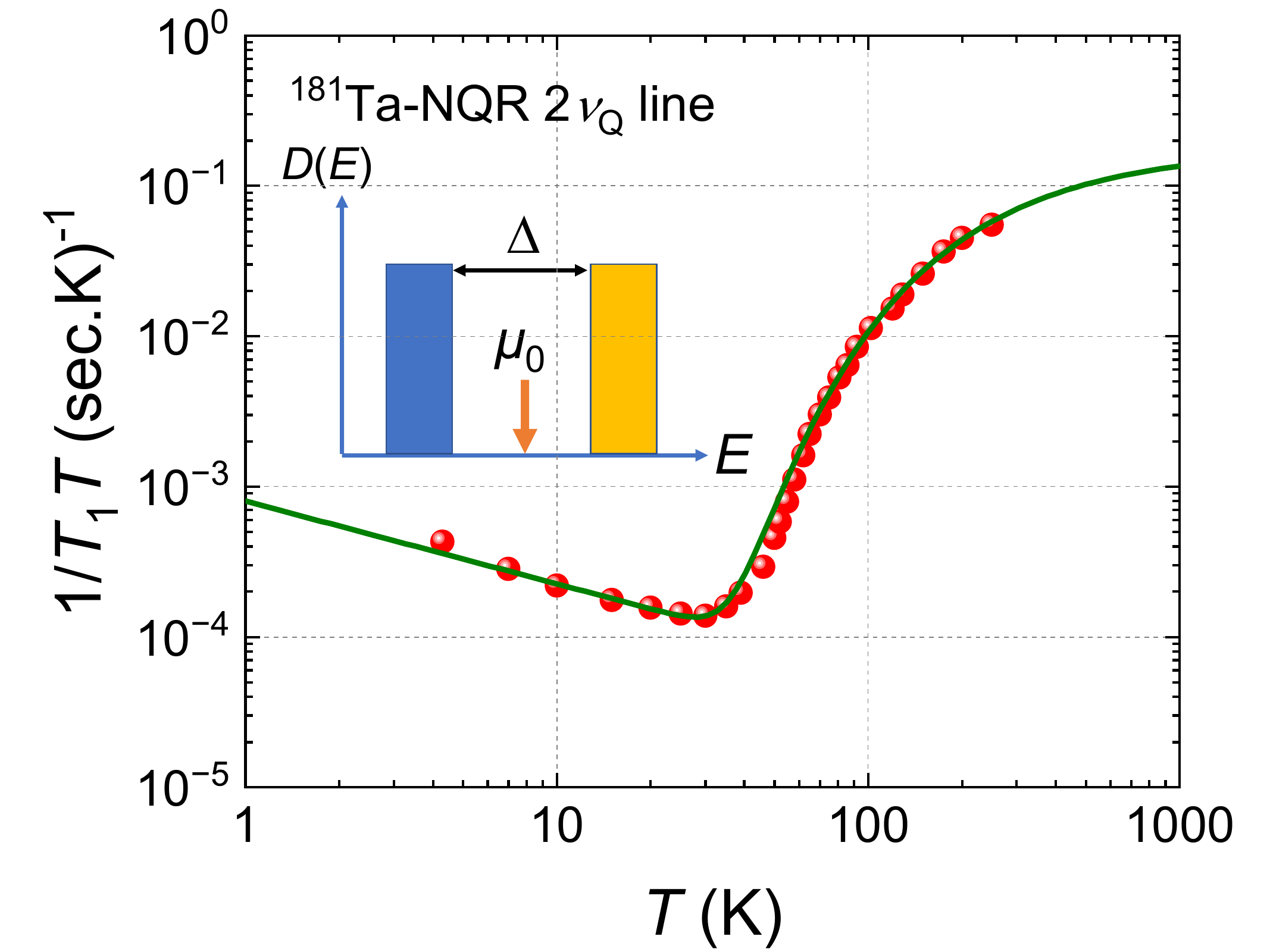}
 \caption{(Color online) Fitting for $1/T_{1} T$ assuming a rectangular in-gap state shown as inset a least-sure fit of the data has adapted to Eq.\,(\ref{eqn:ingap}). The result is shown by a solid curve with low-temperature power law exponent $\beta = 0.55$ and high-temperature activation energy of $\Delta = 283$ K (24.4\,meV).}
 \label{fig:ingap}
\end{figure}

Simple $D(E)$ calculations predict a fairly high energy scale for the excitation of the valence band. To reconcile this with the experimental data, here we assume the existence of rather narrow bands crossing the Fermi energy shown inset of Fig.\,\ref{fig:ingap}. Following the common phenomenological treatment of $1/T_{1} T$ at the high-temperature region, an activated-type temperature dependence of $1/T_{1} T$ has been assumed. Including the low temperature upturn the data have been fitted to the following empirical form,
\begin{equation}
 \frac{1}{T_{1} T} = \alpha T^{-\beta} + \left(\frac{1}{T_{1} T}\right)_{0} \exp \left(- \frac{\Delta}{k_{\rm B} T}\right),
 \label{eqn:ingap}
\end{equation}
where the first term is associated with the local moment type fluctuations of the in-gap state, and the second term is due to an activation process in high temperatures with the energy of $\Delta$. The solid line is a least-squares fit of the data to Eq.\,(\ref{eqn:ingap}). We found $\alpha = 8 \times 10^{-4}$ sec.\,K and $\beta = 0.55$ for the first term, and ($1/T_{1} T)_{0} = 0.18$ sec.\,K and $\Delta/k_{\rm B} = 283$ K (24.4\,meV) for the second term. The energy scale in the activation process found in TaAs is nearly one order of magnitude larger than those observed in similar materials, SmB${_6}$ ($\Delta = 4.3$ meV) \cite{SmB6}, FeGa$_{3}$ ($\Delta = 1.1$ meV) \cite{FeGa3}, and PuB$_{4}$ ($\Delta = 1.8$ meV) \cite{PuB4}. There also exists a common feature of $1/T_{1} T$ in low-temperature region, where the exponential decrease of $1/T_{1} T$ with decreasing temperature crosses over to other excitations which give rise to an upturn of $1/T_{1} T$. Within the present model, the low-temperature behavior must be due to the local moment-type fluctuations in the occupied narrow band. If this is the case, $1/T_{1} T$ should be treated by the exchange narrowed theory and $\beta$ should be one. The fit value, $\beta = 0.55$, may indicate that the assumed in-gap state is rather spatially extended so that the exchange narrowed theory may not be applicable. The origin of the in-gap state is not clear, but it may be associated with Anderson localization \cite{Anderson} or impurity states.

\subsubsection*{Case-3: $1/T_{1} T (T)$ from Weyl fermion excitations}
\begin{figure}[htbp]
 \centering
 \includegraphics[width=1.00\linewidth]{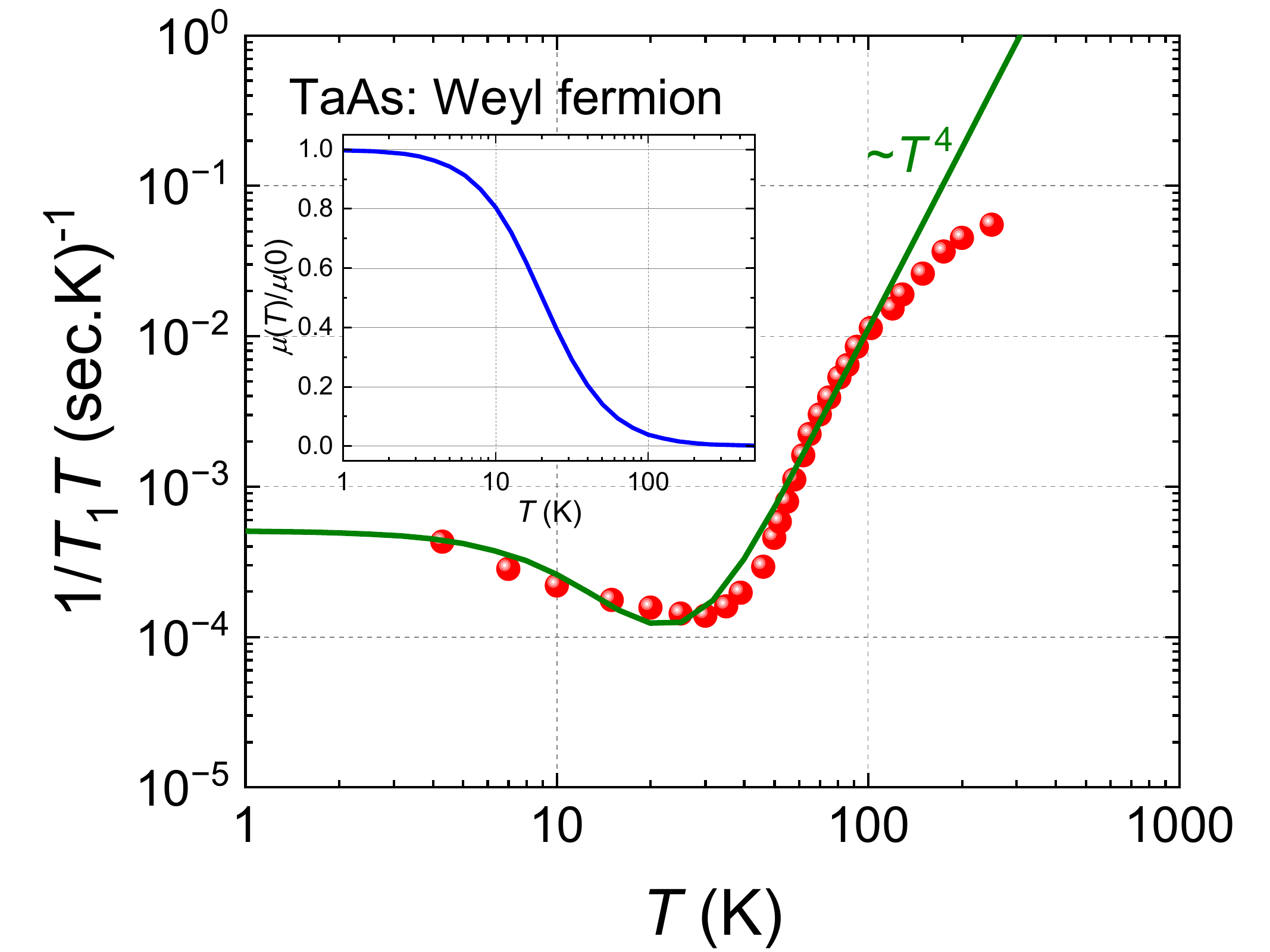}
 \caption{(Color online) Experimental temperature dependence of $1/T_{1} T$ are fitted to theoretical Weyl fermion excitations by a solid curve in TaAs. The normalized temperature dependence of chemical potential, $\mu(T)/\mu(0)$, is shown in the inset.}
 \label{fig:Weyl}
\end{figure}

The first successful observation of the Weyl fermion excitations in topological materials has been achieved in the temperature dependence of $1/T_{1} T$ in TaP where the $T^{2}$ power law dependence has observed in high temperatures \cite{TaPNQR}. This $T^{2}$ dependence was interpreted as competing relaxation channels between the spin and orbital. The spin channel is due to the Weyl fermion excitations associated with the linear dispersion around the Weyl points and has $T^{4}$ dependence of $1/T_{1} T$. The orbital channel is the relaxation process associated with the fluctuations of the orbital hyperfine field which leads to the $T^{-2}$ dependence \cite{Aorb2016,Aorb2019}. In high temperatures, both contributions are equally acting in TaP, then the $T^{2}$ dependence has been observed which is associated with the excitation of Weyl points located 13\,meV above Fermi level. The same scenario was applied to the $T^{2}$ dependence observed in $^{75}$As-NQR measurements in TaAs \cite{TaAsZheng}. However, our $^{181}$Ta-NQR measurements in TaAs revealed $T^{4}$ dependence, meaning the orbital contribution is negligibly small. Following the previous calculation for TaP, $1/T_{1} T$ from the spin channel for TaAs has been calculated using,
\begin{align}
\frac{1}{T_{1} T} & = \alpha \left[4 \mu (T)^{4} + 8 \pi^{2} \mu (T)^{2} \, T^{2} + (28 \pi^{4}/15) \, T^{4}\right], \nonumber \\
\mu (T) & = \frac{\mu(0)}{1 + c [T/\mu(0)]^{2}},
\end{align}
where $\mu (T)$ is the temperature-dependent chemical potential \cite{Aorb2019} in the unit of K. Then we have obtained a reasonably good fit to the data with $\alpha = 6.14 \times 10^{-13}$ sec.$^{-1}$ K$^{-5}$, $\mu(0) = 120$ K, $c = 35$ as shown by a solid curve in Fig.\,\ref{fig:Weyl}. It should be noted that the deviation above $\sim$100\,K may be due to a cutoff effect of the Weyl fermion excitation toward the Korringa process. We also note that we currently have no explanation as to why the orbital relaxation channel is not visible compared with the case of TaP.

Based on the given information, it is difficult to draw a definitive conclusion about the temperature dependence of $1/T_{1} T$. While Case 2 appears to be the most plausible scenario, the lack of information regarding the hyperfine coupling constants of both spin and charge relaxation channels prevents us from making a conclusive statement.

\subsection{Nuclear spin-spin relaxation}
The nuclear spin-spin (transverse) relaxation time, $T_{2}$, was obtained by measuring the spin-echo amplitude, $E$, as a function of the time $t$ between the first exciting pulse and the spin-echo position. The amplitude $E(t)$ can generally be expressed as,
\begin{equation}
 E(t) = E_{0} {\rm e}^{- \Delta^{2} t^{2}} [c_{0} + c_{1} \cos(J t + \phi) {\rm e}^{-t/\tau_{\rm c}}],
 \label{eqn:T2}
\end{equation}
where $\Delta$ is the second moment of the direct nuclear spin-spin interaction due to the classical dipolar coupling in the first place, $c_{0}$ and $c_{1}$ are constants, and the cosine term is the oscillatory term due to the indirect coupling or nuclear quadrupole coupling with their characteristic decay constant, $\tau_{\rm c}$. The spin-echo modulation is well known for the case that the interaction is given by the formula $J (\vec{I}_{i} \cdot \vec{I}_{j}) = J/2 \, (I_{i +} I_{j -} + I_{i -} I_{j +}) + J I_{z}^{2}$. Here, the $I_{z}^2$ term is responsible to the oscillation because this term is invariant for the refocusing pulse ($\pi$-rotation in the rotating frame) making an oscillatory behavior of the formation of spin-echo as a function of $t$. The clear evidences for this oscillated spin-echo decay have been documented for the nuclear quadrupole interaction \cite{SEOM,VO2} and the indirect nuclear spin-spin coupling via conduction electrons (the Ruderman-Kittel interaction) \cite{PtSEO}. The direct nuclear spin-spin coupling includes the same term, but the coupling constant $J$ is so small (the oscillation has a range of several milliseconds at least) that one could not see this effect except for a special case. It should be noted that the direct nuclear spin-spin interaction is easily detuned by inhomogeneous broadening (either external field, sample inhomogeneity, or both) making the decay longer with an exponential function, $E(t) = E_{0} \exp(- \alpha t)$.

\begin{figure}[htbp]
 \centering
 \includegraphics[width=1.00\linewidth]{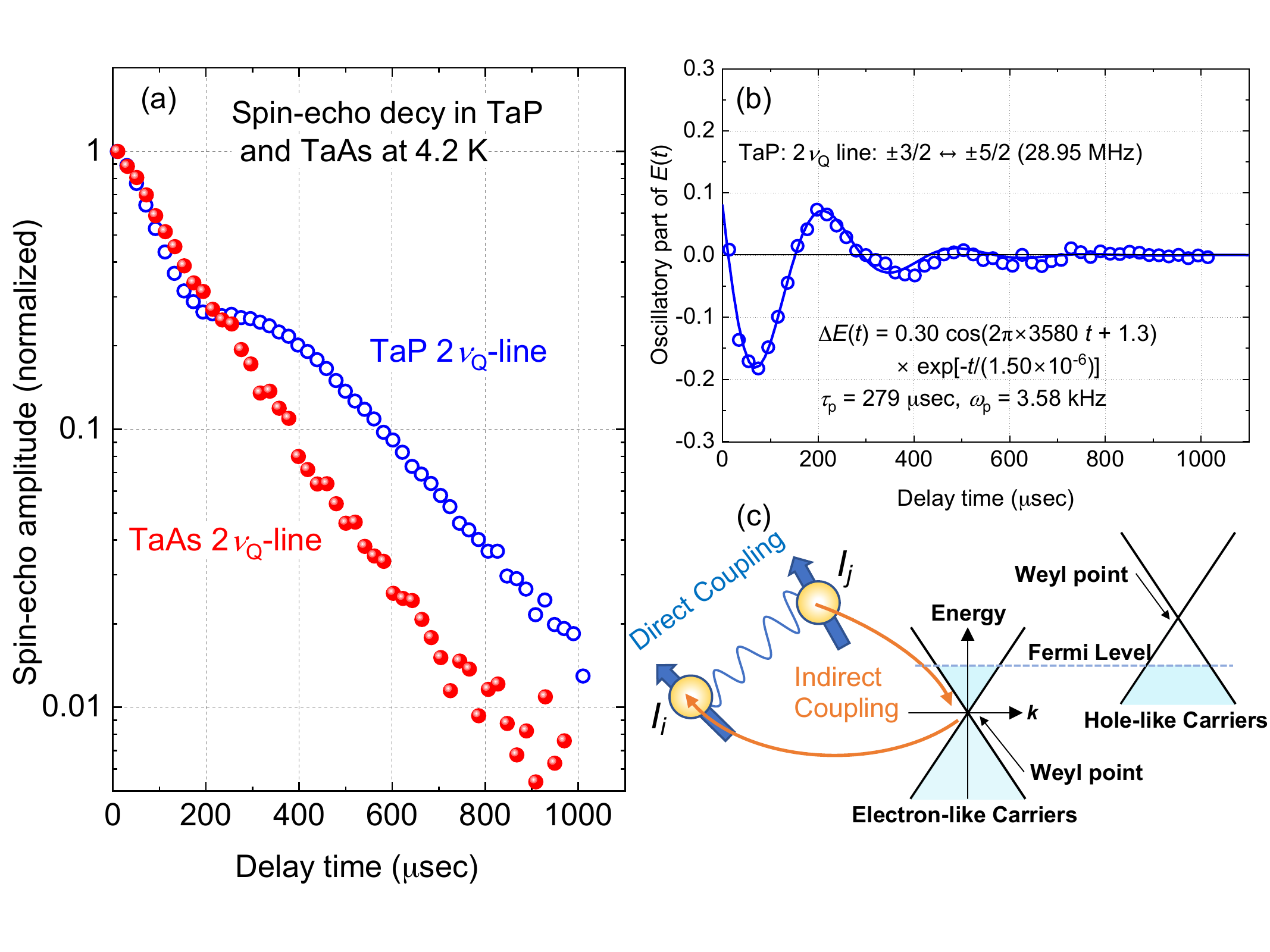}
 \caption{(Color online) (a) Typical spin echo decay for $2 \nu_{\rm Q}$ lines in TaAs (filled circles) and TaP (open circles) at 4.2\,K. A strong sin-echo modulation has been observed in TaP while in TaAs spin-echo decays exponentially without any modulation. The oscillatory part of the spin-echo decay in TaP (open circles) is shown in (b) with the data fit to Eq.\,(\ref{eqn:T2}) (solid curve). A cartoon of an indirect nuclear spin-spin coupling via virtual excitation of Weyl fermions is illustrated in (c).}
 \label{fig:T2}
\end{figure}

The experimental spin-echo decay curves taken at $2 \nu_{\rm Q}$ line of TaAs and TaP are shown in Fig.\,\ref{fig:T2}(a), where spin-echo decays basically exponential for both compounds, but the oscillation was seen only for TaP. As shown in Fig.\,\ref{fig:T2}(b), the oscillatory part of the decay in TaP can be fitted to $\Delta E(t) = c_{1} \cos (\omega_{\rm p} t + \phi) \exp (-t/\tau_{\rm c})$ with $c_{1} = 0.30$, $\omega_{\rm p}/2 \pi = 3.58$\,kHz, and $\tau_{\rm c}$ = 150\,$\mu$sec. with $\omega_{\rm p}$ is an energy scale of indirect coupling. If the oscillation is caused by an indirect nuclear spin-spin coupling via virtual excitation of Weyl fermions as illustrated in Fig.\,\ref{fig:T2}(c), the absence of oscillation in TaAs indicates an absence of Weyl fermion excitations. This may be consistent with the $1/T_{1} T$ behavior discussed in Case 2.

\section{Concluding Remarks}
We presented an extended comparative microscopic study of one of the typical Weyl semimetals, TaAs, beyond previous work on TaP, utilizing the $^{181}$Ta-NQR technique. The experimental results are contrasted between the above two monopnictides. The NQR parameters, $\nu_{\rm Q}$ and $\eta$, are in good agreement with the ab initio calculations for both compounds. However, their temperature dependence above approximately 100\,K shows distinct characteristics, in the sense that $\nu_{\rm Q} (T)$ deviates considerably from calculated values using simple thermal expansion. This discrepancy is likely due to a manifestation of the change of the electronic structure above 100\,K.

Likewise, nuclear spin-lattice relaxation rate $1/T_{1} T$ and nuclear spin-echo decay have great contrast between TaP and TaAs. In TaP, $1/T_{1} T$ are well documented by the Weyl fermion excitations with a temperature-dependent orbital hyperfine interaction. However, in TaAs, we observed a $T^{4}$ power law dependence of $1/T_{1} T$, which could potentially be associated with the linear dispersion of the Weyl fermions within a certain temperature range. Despite this observation, we were unable to draw a conclusive picture for $1/T_{1} T$. The lack of information regarding the hyperfine coupling constants of both spin and charge relaxation channels prevents us from making a conclusive statement.

The work on TaAs shows that there are still many open questions in the field of Weyl semimetals, and this is even more true when trying to understand local measurement methods such as the $^{181}$Ta-NQR. There is an urgent need to use more local methods like NQR and NMR but also muon spin spectroscopy ($\mu$SR) or ESR to fully understand electronic excitations near the Fermi level in detail.

\section*{Acknowledgements}
We thank G. Auffermann, U. Burkhardt, and V. {S\"{u}\ss} for help with the synthesis and characterization of the TaAs crystals.
B. D. was supported by the Ministry of Culture and Innovation and the National Research, Development and Innovation Office within the Quantum Information National Laboratory of Hungary (Grant No. 2022-2.1.1-NL-2022-00004) K134437, K142179 and by a grant of the Ministry of Research, Innovation and Digitization, CNCS/CCCDI-UEFISCDI, under projects number PN-III-P4-ID-PCE-2020-0277.
We thank U. Nitzsche (IFW Dresden) for technical support.
We thank the ESRF (ID22) for providing beamtime.


\begin{thebibliography}{99}
\bibitem{Franz} M. Franz and L. Molenkamp (Eds.), Topological Insulators, Contemporary Concept of Condensed Materials Science Vol. 6, Elsevier 2013, ISBN 978-0-444-63314-9.
\bibitem{mobility2015} C. Shekhar, A. K. Nayak, Y. Sun, M. Schmidt, M. Nicklas, I. Leermakers, U. Zeitler, Y. Skourski, J. Wosnitza, Z. Liu, Y. Chen, W. Schnelle, H. Borrmann, Y. Grin, C. Felser, and B. Yan, Nat. Phys. \textbf{11}, 645 (2015).
\bibitem{SFA} X. Wan, A. M. Turner, A. Vishwanath, and S. Y. Savrasov, Phys. Rev. B \textbf{83}, 205101 (2011).
\bibitem{CME} A. A. Zyuzin, A. A. Burkov, Phys. Rev. B \textbf{86}, 115133 (2012).
\bibitem{TPSS} S.-M. Huang, S.-Y. Xu, I. Belopolski, C.-C. Lee, G. Chang, B. Wang, N. Alidoust, G. Bian, M. Neupane, C. Zhang, S. Jia, A. Bansil, H. Lin, and M. Zahid Hasan, Nat. Commun. \textbf{1}, 8373 (2015).
\bibitem{CANMR} X. Huang, L. Zhao, Y. Long, P. Wang, D. Chen, Z. Yang, H. Liang, M. Xue, H. Weng, Z. Fang, X. Dai, and G. Chen, Phys. Rev. X \textbf{5}, 031023 (2015).
\bibitem{optic} B. Xu, Y. M. Dai, L. X. Zhao, K. Wang, R. Yang, W. Zhang, J. Y. Liu, H. Xiao, G. F. Chen, A. J. Taylor, D. A. Yarotski, R. P. Prasankumar, and X. G. Qiu, Phys. Rev. B \textbf{93}, 121110(R) (2016).
\bibitem{ABJ2016} C.-L. Zhang, S.-Y. Xu, I. Belopolski, Z. Yuan, Z. Lin, B. Tong, G. Bian, N. Alidoust, C.-C. Lee, S.-M. Huang, T.-R. Chang, G. Chang, C.-H. Hsu, H.-T. Jeng, M. Neupane, D. S. Sanchez, H. Zheng, J. Wang, H. Lin, C. Zhang, H.-Z. Lu, S.-Q. Shen, T. Neupert, M. Zahid Hasan, and S. Jia, Nat. Commun. \textbf{7}, 10735-1–9 (2015).
\bibitem{TaPNQR} H. Yasuoka, T. Kubo, Y. Kishimoto, D. Kasinathan, M. Schmidt, B. Yan, Y. Zhang, H. Tou, C. Felser, A. P. Mackenzie, and M. Baenitz, Phys. Rev. Lett. \textbf{118}, 236403 (2017).
\bibitem{Aorb2016} Z. Okv\'{a}tovity, F. Simon, and B. D\'{o}ra, Phys. Rev. B \textbf{94}, 245141 (2016).
\bibitem{Aorb2019} Z. Okv\'{a}tovity, H. Yasuoka, M. Baenitz, F. Simon, and B. D\'{o}ra, Phys. Rev. B \textbf{99}, 115107 (2019).
\bibitem{TaAsQO} F. Arnold, M. Naumann, S.-C. Wu, Y. Sun, M. Schmidt, H. Borrmann, C. Felser, B. Yan, and E. Hassinger, Phys. Rev. Lett. \textbf{117}, 146401 (2016).
\bibitem{TaPQO} F. Arnold, C. Shekhar, S.-C. Wu, Y. Sun, R. D. dos Reis, N. Kumar, M. Naumann, M. O. Ajeesh, M. Schmidt, A. G. Grushin, J. H. Bardarson, M. Baenitz, D. Sokolov, H. Borrmann, M. Nicklas, C. Felser, E. Hassinger, and B. Yan, Nat. Commun. \textbf{7}, 11615 (2016).
\bibitem{Slichter} C. P. Slichter, Principles of Magnetic Resonance 3rd. Ed., Springer -Verlag 1990, ISBN: 978-3-662-09441-9.
\bibitem{Progr} V. F. Mitrovi\'{c}, E. E. Sigmund, and W. P. Halperin, Phys. Rev. B \textbf{64}, 024520 (2001).
\bibitem{etaT1} J. Chepin, and J. H. Ross Jr., J. Phys.: Condens. Matter, 3 8103 (1991).
\bibitem{FPLO} K. Koepernik, and H. Eschrig, Phys. Rev. B \textbf{59}, 1743 (1999).
\bibitem{PW} J. P. Perdew, and Y. Wang, Phys. Rev. B \textbf{45}, 13244 (1992).
\bibitem{wannier90} A. A. Mosto, J. R. Yates, Y. S. Lee, I. Souza, D. Vanderbilt, and N. Marzari, Comput. Phys. Commun. \textbf{178}, 685 (2008).
\bibitem{TaAsZheng} C. G. Wang, Y. Honjo, L. X. Zhao, G. F. Chen, K. Matano, R. Zhou, and G.-q. Zheng, Phys. Rev. B \textbf{101}, 241110(R) (2020).
\bibitem{Abragam} A. Abragam, The Principles of nuclear magnetism, Oxford University Press 1983, ISBN: 978-0-198-52014-6.
\bibitem{SmB6} M. Takigawa, H. Yasuoka, Y. Kitaoka, T. Tanaka, H. Nozaki, and Y. Ishizawa, J. Phys. Soc. Jpn. \textbf{50}, 2525 (1981).
\bibitem{FeGa3} A. A. Gippius, V. Yu. Verchenko, A. V. Tkachev, N. E. Gervits, C. S. Lue, A. A. Tsirlin, N. Büttgen, W. Krätschmer, M. Baenitz, M. Shatruk, and A. V. Shevelkov, Phys. Rev. B \textbf{89}, 104426 (2014).
\bibitem{PuB4} A. P. Dioguardi, H. Yasuoka, S. M. Thomas, H. Sakai, S. K. Cary, S. A. Kozimor, T. E. Albrecht-Schmitt, H. C. Choi, J.-X. Zhu, J. D. Thompson, E. D. Bauer, and F. Ronning, Phys. Rev B \textbf{99}, 035104 (2019).
\bibitem{Anderson} S. Takagi, H. Yasuoka, S. Ogawa, and J. H. Wernick, J. Phys. Soc. Jpn. \textbf{50}, 2539 (1981).
\bibitem{SEOM} H. Abe, H. Yasuoka, and A. Hirai, J. Phys. Soc. Jpn. \textbf{21}, 77 (1966).
\bibitem{VO2} H. Yasuoka, H. Nishihara, Y. Nakamura, and J. P. Remeika, Phys. Lett. \textbf{37A}, 299 (1971).
\bibitem{PtSEO} C. Froidevaux, and M. Weger, Phys. Rev. Lett. \textbf{12}, 123 (1964).
\end{thebibliography}
\end{document}